\input phyzzx.TEX
\tolerance=1000
\voffset=-0.0cm
\hoffset=0.7cm
\sequentialequations
\def\rl{\rightline}

\def\t1{{\tilde 1}}

\def\t{\theta}

\REF{\BEK}{J. Bekenstein, Lett. Nuov. Cimento {\bf 4} (1972) 737; Phys Rev. {\bf D7} (1973) 2333; Phys. Rev. {\bf D9} (1974) 3292.}
\REF{\HAW}{S. Hawking, Nature {\bf 248} (1974) 30; Comm. Math. Phys. {\bf 43} (1975) 199.}
\REF{\STR}{see for example, A. W. Peet, [arXiv:hep-th/0008241]; J. R. David, G. Mandal and S. R. Wadia, Phys. Rep. {\bf 369} (2002) 549, [arXiv:hep-th/0203048] and references therein.} 
\REF{\GIB}{G. W. Gibbons and S. W. Hawking, Phys. Rev. {\bf D15} (1977) 2752.}
\REF{\LEN}{L. Susskind, [arXiv:hep-th/9309145].}
\REF{\CON}{C. Callan and F. Wilczek, Phys. Lett. {\bf B333} (1994) 55, [arXiv:hep-th/9401072 ].}
\REF{\WAL}{R. M. Wald, Phys. Rev. {\bf D48} (1993) 3427, [arXiv:gr-gc/9307038]; V. Iyer and R. M. Wald, Phys. Rev. {\bf D50} (1994) 846, [arXiv:gr-qc/9403028]; Phys. Rev. {\bf D52} (1995) 4430, [arXiv:gr-qc/9503052].}
\REF{\SBH}{E. Halyo, A. Rajaraman and L. Susskind, Phys. Lett. {\bf B392} (1997) 319, [arXiv:hep-th/9605112].}
\REF{\HRS}{E. Halyo, B. Kol, A. Rajaraman and L. Susskind, Phys. Lett. {\bf B401} (1997) 15, [arXiv:hep-th/9609075].}
\REF{\EDI}{E. Halyo, Int. Journ. Mod. Phys. {\bf A14} (1999) 3831, [arXiv:hep-th/9610068]; Mod. Phys. Lett. {\bf A13} (1998), [arXiv:hep-th/9611175].}
\REF{\MED}{A. J. M. Medved, [arXiv:hep-th/0201215].}
\REF{\DES}{E. Halyo, [arXiv:hep-th/0107169].}
\REF{\UNI}{E. Halyo, JHEP {\bf 0112} (2001) 005, [arXiv:hep-th/0108167].}
\REF{\JAC}{T. Jacobson, G. Kang and R. C. Myers, Phys. Rev. {\bf D49} (1994) 6587, [arXiv:gr-qc/9312023]; [arXiv:gr-qc/9502009].}
\REF{\LOV}{D. Lovelock, Journ. Math. Phys. {\bf(12)} (3) (1971) 498.}
\REF{\GB}{T. Clunan, S. F. Ross and D. J. Smith, Class. Quant. Grav. {\bf 21} (2004) 3447, [arXiv:gr-qc/0402044].}
\REF{\GOL}{H. Goldstein, "Classical Mechanics" Addison Wesley  Publishing Co. (1980) pp. 457-462.}
\REF{\HOL}{G. 't Hooft, [arXiv:gr-qc/9310026]; L. Susskind, J. Math. Phys. {\bf 36} (1995) 6377, [arXiv:hep-th/9409089]; R. Bousso, Rev. Mod. Phys. {\bf 74} (2002) 825, [arXiv:hep-th/0203101].}
\REF{\ADS}{J. Maldacena, Adv. Theor. Math. Phys. {\bf 2} (1998) 231, [arXiv:hep-th/9711200]; S. Gubser, I. Klebanov and A. Polyakov, Phys. Lett. {\bf B428} (1998) 105,
[arXiv:hep-th/9802109]; E. Witten, Adv. Theor. Math. Phys. {\bf 2} (1998) 253, [arXiv:hep-th/9802150].}
\REF{\SON}{E. Halyo, in preparation.}

\singlespace
\rl{SU-ITP-14/03}
\pagenumber=0
\normalspace
\medskip
\bigskip
\titlestyle{\bf{Rindler Energy is Wald Entropy}}
\smallskip
\author{ Edi Halyo{\footnote*{e--mail address: halyo@stanford.edu}}}
\smallskip
\centerline {Department of Physics} 
\centerline{Stanford University} 
\centerline {Stanford, CA 94305}
\smallskip
\vskip 2 cm
\titlestyle{\bf ABSTRACT}
We show that, in any theory of gravity, the entropy of any nonextreme black hole is given by $2 \pi E_R$ where $E_R$ is the dimensionless Rindler energy. Separately, we show that $E_R$ is exactly Wald's Noether charge and therefore this entropy is identical to Wald entropy. However, it is off--shell and derived solely from the time evolution of the black hole.
We examine Gauss--Bonnet black holes as an example and speculate on the degrees of freedom that $E_R$ counts.

\singlespace
\vskip 0.5cm
\endpage
\normalspace

\centerline{\bf 1. Introduction}
\medskip

The correct quantum theory of gravity is expected to explain the origin of the 
Bekenstein--Hawking entropy of black holes[\BEK,\HAW] in terms of the fundamental gravitational degrees of freedom.
For example, in string theory, the degrees of freedom that give rise to black hole entropy for a large class of BPS and near--BPS 
black holes[\STR] are well--known. However, the same cannot be said for black objects far from extremality. 
The entropy of these black holes can be computed only at the macroscopic level, i.e. in General Relativity,
by a number of methods such as those that use the Euclidean gravitational action[\GIB], the conical deficit angle[\LEN,\CON] and Wald's Noether charge[\WAL]. On the other hand, any quantum
theory of gravity, at energies much lower than the Planck scale, reduces to General Relativity modified by additional effective terms
which are inversely proportional powers of the Planck mass. Thus, a method for computing black hole entropy, if it has any chance of
leading to a deeper understanding of the fundamental degrees of freedom, should also apply
to these generalized theories of gravity. The most popular method, in this respect, is the one that uses Wald's Noether charge (even though other methods, such as the deficit angle method also work). Unfortunately, since these methods only use the classical geometry or the metric the 
microscopic degrees of freedom that they count are obscure. Therefore, it is important to find 
new methods for computing black hole entropy in generalized theories of gravity with
the hope that they may shed light on the fundamental degrees of freedom of quantum gravity.

There are two properties common to all conventional methods for computing black hole entropy[\GIB,\LEN,\CON,\WAL]. First,
entropy is given by a surface integral over the black hole horizon. Thus, even though entropy is
not proportional to horizon area in generalized theories of gravity, these are presumably still holographic since 
only the horizon degrees of freedom seem to contribute to the entropy. Thus, it would be interesting to see if there is a way to compute black hole entropy that is not related to the horizon surface and therefore not manifestly holographic.
Second, entropy is an on--shell quantity; it is obtained by assuming that field equations are satisfied. In particular, one needs to know the Lagrangian of the gravitational theory in order to compute the entropy. However, the other thermodynamic quantity, Hawking temperature is off-shell, i.e. it is computed directly from the
metric which may or may not solve the field equations. Moreover, one does not even need to know the gravitational Lagrangian. 
A method for computing entropy off--shell would allow us to consider black hole thermodynamics off-shell.

A method for computing black hole entropy that differs from the others with respect to the properties mentioned above uses the dimensionless Rindler energy, $E_R$, of a black object[\LEN]. It is obtained by taking the near horizon limit of the black hole metric which is Rindler space and then rescaling Rindler time to make it dimensionless. $E_R$ is canonically conjugate to this dimensionless Rindler time. 
It has been known for a while that, in General Relativity, black hole entropy is given by $S=2 \pi E_R$. 
{\footnote1{It is easy to see this
by using the dimensionless Rindler temperature $T_R=1/2 \pi$ in the First Law of Thermodynamics.}}
This entropy has been explicitly computed, in any dimension, for
Schwarzschild black holes[\SBH,\HRS], near--extreme charged black holes[\EDI], black branes[\UNI], black holes in anti de Sitter space[\MED] and de Sitter space[\DES].
In this paper, we show that $E_R$, by construction,
satisfies the First Law of Thermodynamics in any theory of gravity. As a result, the entropy of any black
hole in any theory of gravity is given by $S=2 \pi E_R$. We also show that $E_R$ is
identical to Wald's Noether charge which is to be expected since they are both charges (or generators) of dimensionless time translations on the horizon.
Therefore, $2 \pi E_R$ is exactly Wald entropy that is often computed in generalized theories of gravity. 

Since $E_R$ is the energy in Rindler space, it is only related to time evolution and not to the black hole horizon. Thus, its  relation to holography is not clear. In addition, $E_R$ is an off-shell quantity obtained directly from the metric without any 
knowledge of the field equations or the underlying gravitational Lagrangian. Clearly, when both entropy and Hawking temperature are computed off-shell and obtained from the time evolution in the near horizon region of the black hole, we get a more unified description of black hole thermodynamics which may lead to a better understanding of these issues.

The fundamental degrees of freedom that $E_R$ counts are obscure since $E_R$ is a classical quantity obtained solely from the metric. In the near horizon region and in terms of dimensionless Euclidean Rindler time, the ($t-r$ part of the) black hole metric is simply the flat metric in polar coordinates. All the information about the black hole is carried by $E_R$. In the absence of any quantum theory that describes
the black hole, we can only speculate on what $E_R$ counts.
Since $E_R$ is conjugate to the dimensionless Euclidean Rindler time, i.e. an angle, it can be seen as the dimensionless frequency of a simple one dimensional, periodic system. Then, $S= 2 \pi E_R=2 \pi \omega$, where $\omega$ is the dimensionless angular frequency,{\footnote2{This is the analog of the (dimensionless) Matsubara frequency in a thermal field theory.}} which is an integer due to quantization along the periodic
(dimensionless Euclidean) time direction. Alternatively, one can consider the action variable of a one dimensional periodic     system, $J=\oint p~ dq$ which is quantized. Using the dimensionless Rindler time and energy for the canonical variables $p$ and $q$ we find that $J=2 \pi E_R$ gives the black hole entropy. In either description, the quantum
mechanics of a periodic system with one degree of freedom seems to give the entropy of the black hole which is a thermal ensemble.
Unfortunately, we understand neither the degrees of freedom $E_R$ counts nor the quantum theory of gravity that describes 
them. It would be interesting to see if these ideas can be realized in a string theory description of black holes{\footnote3{For an early attempt in string theory see ref. [\LEN] and the comments at the end of section 5.}}. 


The paper is organized as follows. In the next section, we describe the dimensionless Rindler energy and show that $S= 2 \pi E_R$
for any black hole in any theory of gravity. In section 3, we show that $E_R$ is exactly Wald's Noether charge and therefore
$2 \pi E_R$ is identical to Wald entropy. In section 4, we show that the entropy of black holes in Gauss--Bonnet gravity 
obtained through $E_R$ agrees with the Wald entropy and discuss a sufficient condition required to have corrections to the area law. Section 5 contains some speculations about the meaning of $E_R$ and the degrees of freedom that $E_R$ counts. Section 6 contains a discussion of our results and our conclusions.

\bigskip
\centerline{\bf 2. Dimensionless Rindler Energy as Entropy}
\medskip

Consider any non--extreme black hole with a generic metric of the form
$$ ds^2=-f(r)~ dt^2+ f(r)^{-1} dr^2+ r^2 d \Omega^2_{D-2} \quad, \eqno(1)$$
in D-dimensions. The horizon is at $r_h$ which is determined by
$f(r_h)=0$. If in addition, $f^{\prime}(r_h) \not =0$, the near horizon region is described by Rindler space. Near the
horizon, $r=r_h +y$ where $y<<r_h$. Thus, to first order,
$f(r)=f(r_h)+f^{\prime}(r_h)y$ and we get the near horizon metric
$$ds^2=-f^{\prime}(r_h)y~ dt^2+(f^{\prime}(r_h)y)^{-1} dy^2+ r_h^2 d \Omega^2_{D-2} \quad. \eqno(2)$$
In terms of the proper radial distance, $\rho$, obtained from $d\rho=dy/\sqrt{f^{\prime}(r_h)y}$ the metric becomes
$$ds^2=-{f^{\prime 2}(r_h) \over 4} \rho^2 dt^2+d \rho^2+ r_h^2 d \Omega^2_{D-2} \quad. \eqno(3)$$
After a rescaling to get the dimensionless Rindler time $\tau_R=(f^{\prime}(r_h)/2)~ t$  we find
$$ds^2=-\rho^2 d \tau_R^2 + d \rho^2 + r_h^2 d \Omega^2_{D-2} \quad, \eqno(4)$$
where the metric in the $\tau_R-\rho$ directions describes Rindler space.

Originally, in Ref.[\LEN] it was shown that the entropy of a black hole is given by the dimensionless Rindler energy $E_R$ conjugate to $\tau_R$. $E_R$ can be calculated using the Poisson bracket
$$1=\{E_R,\tau_R\}=\left({\partial E_R \over \partial M}{\partial \tau_R \over \partial t}-{\partial E_R \over \partial t}
{\partial \tau_R \over \partial M} \right) \quad, \eqno(5)$$
where $M$ is the mass of the black object conjugate to $t$. This transition from the canonical variables $t,M$ to $\tau_R,E_R$
is a canonical transformation. Taking $E_R$ to be time independent which is a good approximation for large enough black holes (with Hawking radiation slow enough to be negligible) we find
$$dE_R={2 \over f^{\prime}(r_h)}~ dM \quad.\eqno(6)$$
The entropy of the black hole is then given by $S=2 \pi E_R$. This method works for all nonextreme black objects in any dimension with Rindler--like near horizon geometries. This result is perhaps not so surprising in light of the fact that the above derivation of $E_R$ is equivalent to the
First Law of Thermodynamics. Using the definition of Hawking temperature obtained from the metric, $T_H=f^{\prime}(r_h)/4 \pi$, eq. (6) can be written as
$$d(2 \pi E_R)={dM \over T_H} \quad. \eqno(7)$$
As a result, $S=2 \pi E_R$ for all black objects in all dimensions. 

The arguments above are completely general and apply beyond General Relativity. Note that
in any theory of gravity, the generic black hole metric in eq.(1) has the same form  with $f(r_h)=0$ and 
$f^{\prime}(r_h) \not=0$ (even though the exact form of the factor $f(r)$ differs among theories). In addition, the definition of Hawking temperature, $T_H=f^{\prime}(r_h)/4 \pi$, is universal.
Therefore, eq. (7) holds in any theory of gravity.
Since, by construction, $E_R$ satisfies the First Law of Thermodynamics we can conclude that
$S=2\pi E_R$ in any theory of gravity.
In the next section we will obtain the same result by showing
that $E_R$ is exactly Wald's Noether charge and therefore $2 \pi E_R$ is identical to Wald entropy.

We note that the calculation the entropy through $E_R$ is not a new 
method since it is completely equivalent to using thermodynamics, i.e. the relation $T(M)$, to obtain the entropy. However, $E_R$
has two properties that are not shared by the other methods for computing entropy. First, $E_R$ is an off--shell quantity that gives the entropy related to metrics that may or may not solve the field equations. In fact, $E_R$ is calculated directly from the metric 
with no knowledge of the underlying gravitational Lagrangian. As far as we know, all other methods of computing entropy
are on--shell and explicitly use the gravitational field equations. Second, $E_R$ is obtained only from the $g_{00}$ component of the metric and is not manifestly holographic. Rather, it shows that entropy is somehow related to time evolution.
In all other methods, entropy is obtained from a surface integral over the horizon making the holographic nature of gravity transparent. In General Relativity this leads to the fact that black hole entropy is the horizon area in Planck units. For more general theories, entropy is not proportional to horizon area but the surface integral suggests that holography is still a valid concept. 

It is interesting to note that just like $E_R$, Hawking temperature is also an off--shell quantity that is obtained from the time evolution in the near horizon region. In any theory of gravity, it is computed directly from the Euclidean black hole metric through the periodicity of the time direction. This means that the $g_{00}$ component of the metric completely determines black hole thermodynamics on or off--shell.

\bigskip
\centerline{\bf 3. Dimensionless Rindler Energy and Wald Entropy}
\medskip

In this section, we show that the dimensionless Rindler energy, $E_R$, is exactly Wald's
(rescaled) Noether charge $\tilde Q$[\WAL] and therefore $2 \pi E_R$ is Wald entropy which gives the black hole entropy
in any theory of gravity.
Consider a generalized gravity theory that is described by a Lagrangian of the form $L=L(\phi_i, \nabla_a \phi_i, g_{ab},R_{abcd})$,
i.e. $L$ is a function of the metric, the Riemann tensor and matter fields $\phi_i$ and includes no more than the second derivatives of the metric and first derivatives of the matter fields. This can be generalized to Lagrangians with any number of derivatives of the metric to any order[\WAL] but for simplicity we consider
the more limited case above. None of our results depends on the details of the gravity Lagrangian.

Under a general variation of a field $\psi$ which may be a matter field or the metric, the variation in the Lagrangian can be
written as[\WAL,\JAC]
$$\delta(\sqrt{-g}L)=\sqrt{-g}E \cdot \delta \psi+ \sqrt{-g} \nabla_a \Theta^a \delta \psi \quad, \eqno(8)$$
where the dot indicates summation over fields and contraction of all indices and $E$ denotes the equations of motion. For a diffeomorphism with Killing vector $\xi^a$
the variation of the field is given by the Lie derivative $\delta \psi={\cal L}_{\xi}\psi$ under which a covariant Lagrangian changes by a total derivative
$$\delta (\sqrt{-g}L)={\cal L}_{\xi}(\sqrt{-g}L)=\sqrt{-g} \nabla_a(\xi^a L) \quad. \eqno(9)$$
As a result, one can define a Noether current
$$J^a=\Theta^a ({\cal L}_{\xi} \psi)-\xi^a L \quad, \eqno(10)$$
which is conserved, i.e. $\nabla_aJ^a=0$ when the equations of motion are satisfied and $E=0$. One can define the antisymmetric Noether potential $Q^{ab}$ that satisfies $J^a=\nabla_b Q^{ab}$ and the Noether charge 
$$\int_{\partial V} d^{D-2}x \sqrt{h}~ Q=\int_{\partial V} d^{D-2}x \sqrt{h} \epsilon_{ab}  Q^{ab} \quad, \eqno(11)$$
where $\partial V$ is the ($D-2$ dimensional) boundary of $V$. $h_{ab}$ and $\epsilon_{ab}$ are the induced metric and binormal form on the boundary respectively.

Consider now any black hole with a bifurcate Killing horizon on which the time like Killing vector $\xi^a$ vanishes. We can rescale
$\xi^a \to \kappa \xi^a$ so that the surface gravity is normalized to unity. As a result, the Noether charge also gets rescaled
$Q \to {\tilde Q}=Q/\kappa$.
Then, the Wald entropy of the black hole is given by[\WAL]
$$S_{Wald}= 2 \pi \int_{\partial V} d^{D-2}x \sqrt{h}~{\tilde Q} = - 2 \pi \int_H d^{D-2}x \sqrt{h} Y^{abcd}{\epsilon}_{ab}{\epsilon}_{cd} \quad, \eqno(12)$$
where $\int \tilde Q$ is the Noether charge of the rescaled Killing field and the integral is taken 
over the bifurcation surface of the
Killing horizon with induced metric $h$. The tensor $Y^{abcd}$ is defined by
$Y^{abcd}= {\partial L / {\partial R_{abcd}}}$ and has all the symmetries of the Riemann tensor. When the Lagrangian, $L$
includes higher order terms in $R$ or terms with higher derivatives, the tensor $Y^{abcd}$ gets modified but the eq. (12) still gives Wald entropy.
This entropy by construction satisfies the First Law of Thermodynamics in all theories of gravity.

It is now obvious that the dimensionless Rindler energy, $E_R$ is exactly Wald's Noether charge since
$$S=2 \pi \int_{\partial V} d^{D-2}x \sqrt{h}~ {\tilde Q}=2 \pi E_R \quad, \eqno(13)$$
in any theory of gravity. In addition, calculations of both $\tilde Q$ and $E_R$ apply to
the same types of gravitational backgrounds. The formula for Wald entropy, eq. (12), applies to black holes with a nondegenarate
bifurcate Killing horizons which are exactly those with near horizon geometries described by Rindler space, i.e. metrics
of the type given in eq. (1).

The equivalence between $E_R$ and $\tilde Q$ can be made more precise.
The energy $E$ of a black hole, related to a time like Killing vector $\xi$, can be written as (in terms of forms for simplicity)[\WAL]
$$E=\int_{\infty} Q(\xi)-\xi \cdot B \quad, \eqno(14)$$
where the integral is taken over a hypersurface at infinity and $B$ is a form whose variation is related to $\Theta$ in eq. (8) by
$$\delta \int_{\infty}\xi \cdot B= \int_{\infty} \xi \cdot \Theta \quad. \eqno(15)$$
If we evaluate the integral in eq. (14) on the horizon, $H$, we get
$$E_{hor}=\int_H Q(\xi) \quad, \eqno(16)$$
since the Killing vector $\xi$ vanishes on the horizon. This is the energy as measured on (or near) the horizon.
If in addition we rescale the Killing vector by $\xi \to \kappa \xi$ so that $Q \to {\tilde Q}=Q/\kappa$, we obtain the dimensionless energy on the horizon
$$E_R=\int_H {\tilde Q}(\xi) \quad, \eqno(17)$$
which is precisely eq. (13). 
This shows that the rescaled Noether charge of the time--like vector $\xi$ is the dimensionless Rindler energy. Since the original work in Ref.[\WAL], it has been noticed that the rescaled Noether charge $\int \tilde Q(\xi)$ is conjugate to time translations (defined by the the time--like Killing vector $\xi$) when the surface
gravity $\kappa =1$. In other words, it is the same as,$E_R$, the energy conjugate to dimensionless time when the temperature is rescaled to be $1/2\pi$.

Even though, $\int \tilde Q$ and $E_R$ are both equal to $S/2 \pi$ and therefore describe generalized black hole entropy, they are
conceptually very different. The Noether charge is a surface integral over the horizon and therefore holographic
even when entropy is not given by the horizon area. On the other hand, the Rindler energy, $E_R$, is not related to the horizon surface but obtained only from the $g_{00}$ component of the metric, i.e. the time evolution in the near horizon region.
The computation of the Noether charge explicitly depends on the field equations as seen from
eq. (8) that determines $\Theta$ or from eq. (12) which contains the tensor $Y^{abcd}$ that is obtained from the the gravitational Lagrangian. On the other hand, $E_R$ is computed off-shell, i.e. directly from the metric which may or may not satisfy the gravitational field equations.  

\bigskip
\centerline{\bf 4. $E_R$ and Gauss--Bonnet Black Holes}
\medskip

Effective (nonrenormalizable) theories of gravity beyond General Relativity are described by actions that include, in addition to the Einstein--Hilbert term, higher order terms with coefficients that are proportional to inverse powers of the Planck mass.
These terms describe the effects of the fundamental theory at low energies and parametrize the deviations from General
Relativity. This is certainly the case in all string theories which are the best candidates for quantum gravity. 

In order to compute black hole entropy through $E_R$ in a generalized theory of gravity, we consider 
Gauss--Bonnet gravity with the Lagrangian[\LOV]
$$\sqrt{-g}L={1 \over {16 \pi G}} \sqrt{-g}(R-2 \Lambda + \alpha(R_{abcd} R^{abcd}-4R_{ab}R^{ab}+R^2) \quad, \eqno(18)$$
where $\alpha$ is the coefficient of the Gauss--Bonnet term and the cosmological constant $\Lambda=-(D-1)(D-2)/2l^2$ is included for generality.
The Lagrangian in eq. (18) is obtained in the low energy limit of heterotic string theory with $\alpha=1/2 \pi \ell_s$ parametrizing the stringy corrections to the Einstein--Hilbert term.
For $D=4$, the Gauss--Bonnet term is topological and does not lead to new physics.
Gauss--Bonnet black hole solutions exist for $D \geq 5$ with the metric[\GB]
$$ds^2=-f(r)~dt^2+f(r)^{-1} dr^2+ r^2 h_{ij} dx^idx^j \quad. \eqno(19)$$
The indices $i,j$ run over the $D-2$ transverse directions with a metric of constant curvature $h_{ij}$ equal to $(D-2)(D-3)k$
where $k=0, \pm 1$. The solution in eq. (19) is defined by{\footnote4{The review of Gauss--Bonnet black holes below closely follows
ref. [\GB].}
$$f(r)=k+{r^2\over {2 \alpha}} \left(1-\sqrt{1+{{64 \pi G \alpha M} \over {(D-2)V_k r^{n-1}}}+{{8 \alpha \Lambda} \over {(D-1)(D-2)}}}
\right)  \quad, \eqno(20)$$
where $V_k$ is the volume of the transverse space and with an abuse of notation $\alpha$ has been rescaled by a factor of $(D-3)(D-4)$.

The radius of the black hole horizon, $r_h$, is determined by $F(r_h)=0$. (If there are two roots to the equation, e.g. for vanishing or negative 
cosmological constant, the largest one gives the location of the horizon.) In terms of $r_h$, the mass of the black hole is
$$M={{(D-2)V_k r_h^{n-3}} \over {16 \pi G}} \left(k+ {\alpha \over r_h^2}+ {r_h^2 \over l^2} \right) \quad. \eqno(21)$$
The Hawking temperature is obtained as usual from the periodicity of the Euclidean time direction
$$T_H={{(D-1)r_h^4+(D-3)k l^2 r_h^2 +(D-5)\alpha k^2 l^2} \over {4 \pi l^2 r_h(r_h^2+2 \alpha k)}} \quad. \eqno(22)$$
Now we can use eq. (6) to obtain $E_R$ which is basically equivalent to using the First Law of Thermodynamics to find the entropy from 
$M$ and $T$:
$$S=2 \pi E_R= \int {dM \over T_H}=\int_0^{~r_h}T_H^{-1} \left( {\partial M \over \partial r_h} \right) dr_h \quad, \eqno(23)$$
which gives
$$S={{r_h^{D-2} V_k} \over {4G}} \left(1+{{2 \alpha k(D-2)} \over {(D-4)r_h^2}}\right) \quad. \eqno(24)$$

Note that the correction to the area law is not only proportional, as expected, to $\alpha$ which parametrizes the 
Gauss--Bonnet terms in the Lagrangian but also to $k$, i.e. the curvature of the horizon. In particular, for $k=0$ there is no correction and the black hole entropy is given by the horizon area even in Gauss--Bonnet gravity. 

The Wald entropy of Gauss--Bonnet black holes has been calculated in ref. [\GB]. Using the Lagrangian in eq. (18) to get $Y^{abcd}$
which is substituted in eq. (12) one finds
$$S_{Wald}={1 \over {4G}} \int_H d^{D-2}x \sqrt{\tilde h} (1+2 \alpha {\tilde R}) \quad, \eqno(25)$$
where $h$ is the induced metric on the horizon $H$ and $\tilde R$ is the Ricci scalar of $\tilde h_{ij}=r_h^2 h_{ij}$
which reproduces the entropy in eq. (24). 

It is important to learn the origin of the above corrections to entropy which are proportional to both $\alpha$ and $k$.
The correction term in eq. (24) can be traced back to the second term in the numerator of $\beta=1/T_H$ in eq. (22). In all cases for which the numerator of $\beta$ contains a single term, black hole entropy is proportional to horizon area.
This, of course happens when $\alpha=0$ and Gauss--Bonnet gravity reduces to General Relativity. We see that even when $\alpha \not =0$, the numerator of $\beta$ contains only one term if $k=0$, i.e. for black holes with flat horizons. Thus, we conclude that a sufficient condition for getting corrections to the area law is for $\beta$ to have more than one term in its numerator. 
In light of this, we would like to find out more about the generic form of the factor $f(r)$ in the black hole metric and
corrections to black hole entropy.


The most general D--dimensional black hole metric in the form of eq. (1) can be written with $f(r)$ given by (with $k=1$)
$$f(r)=\left(1 - {{\alpha M} \over r^{D-3}}\right)F(r) +g(r) \quad. \eqno(26)$$
where $\alpha=16 \pi/(D-2)V_{(D-2)}$ (not to be confused with the coefficient of Gauss--Bonnet gravity). In General Relativity either $F(r)=1$ or $g(r)=0$ but in more general theories we have
$F(r) \not=1$ and $g(r) \not=0$. We want to show that, In General Relativity, it is the special form of the metric 
with $F(r)=1$ or $g(r)=0$ that leads to the area law. 

Consider first a metric of the form in eq. (1) with $F(r)=1$ but $g(r) \not =0$.
Using $\beta=4 \pi/ f^{\prime}(r_h)$ we obtain
$$\beta={{4 \pi r_h} \over {(D-3)(1+g(r_h))+ r_h g^{\prime}(r_h)}} \quad. \eqno(27)$$
It is easy to show that using 
$${\partial M \over \partial r_h} ={{(D-2) V_k} \over {16 \pi G}}[(D-3)r_h^{D-4}(1+g(r_h))+r_h^{D-3}g^{\prime}(r_h)] \eqno(28)$$
in eq. (23) we get 
$$E_R= {{(D-2) V_k} \over {8 \pi G}}  \int_0^{r_h} r_h^{D-3} dr_h \quad, \eqno(29)$$ 
and $S=2 \pi E_R$ gives the area law as expected. One can easily repeat the same exercise with $F(r) \not=1$ and $g(r)=0$ which 
again leads to the area law. Thus, the special form of $f(r)$ in eq. (26) with either $F(r)=1$ or
$g(r)=0$ which is satisfied in General Relativity is a sufficient condition for the area law of black hole entropy.

Metrics with both $F(r) \not=1$ and $g(r) \not=0$ do not arise in General Relativity so it is interesting to find out whether they
also lead to the area law. In this case, we find
$$\beta={{4 \pi r_h F(r_h)} \over {[(D-3)(F^2(r_h)+F(r_h)g(r_h))-g^{\prime}(r_h) F(r_h) r_h-g(r_h)F^{\prime}(r_h) r_h]}} \eqno(30)$$
and also
$${\partial M \over \partial r_h} ={(r_h^{D-4} / {\alpha F^2(r_h))} \over {[(D-3)(F^2(r_h)+F(r_h)g(r_h))-r_h(g^{\prime}(r_h) F(r_h)-g(r_h)F^{\prime}(r_h))]}} \quad . \eqno(31)$$
As a result, the $E_R$ becomes
$$E_R= {{(D-2) V_k} \over {8 \pi G}}  \int_0^{r_h} {r_h^{D-3} \over F(r_h)} dr_h \quad . \eqno(32)$$ 
Since $F(r) \not=1$ we find deviations from the area law and entropy is not proportional to the horizon area. 
In Gauss--Bonnet gravity, $f(r)$ in eq. (20) is more complicated than the ansatz we considered in eq. (26). Nevertheless it does not satisfy the conditions mentioned above and as expected leads to eq. (24) which deviates from the area law.

\endpage
\bigskip
\centerline{\bf 5. Speculations on $E_R$}
\medskip

Above we saw that $2 \pi E_R$ gives the correct entropy for all black holes in any theory of gravity. Since we only used the black hole metric, at this level of analysis, it is impossible to describe the microscopic, fundamental degrees of freedom that $E_R$ counts. Clearly, $E_R$ which is the dimensionless Rindler energy is related to the time evolution
in the near horizon region of the black hole space--time and not to the horizon surface. Thus, we do not expect the degrees of freedom that $E_R$ counts to be directly related Planck area cells on the horizon. We can nevertheless try to gain some insight into the microscopic nature of $E_R$. In this section, we consider some speculative ideas on what $E_R$ counts which, hopefully, may lead to a better understanding of black hole entropy.

Consider the Rindler metric in eq. (4) in Euclidean time $\tau_{RE}=i \tau_R$
$$ds^2=\rho^2 d \tau_{RE}^2 + d \rho^2 + r_h^2 d \Omega^2_{D-2} \quad, \eqno(33)$$
where the metric in the $\tau_{RE}-\rho$ directions is simply the flat metric in polar coordinates. This metric is the same for all nonextreme black holes in any theory of gravity. All the information about the black hole now resides in (Euclidean) $E_R$ which is canonically conjugate to $\tau_{RE}$ by eq. (5). In Lorenztian signature, the Rindler coordinates that describe Rindler observers
at fixed $\rho$ are related to the flat coordinates, $X$ and $T$, that describe freely falling observers by
$$T=\rho~ sinh~ \tau_R \qquad  X=\rho~ cosh~ \tau_R \quad, \eqno(34)$$
where the usual factor of the surface gravity is missing since $\kappa=1$ when we use the dimensionless Rindler time $\tau_R$.
These parametrize what is usually called the right Rindler wedge with $X>|T|$ and $X>0$. 
In the Euclidean space--time that is described by the metric in eq. (33), the relations in eq. (34) become
$$T=\rho~ sin~ \tau_{RE} \qquad X=\rho~ cos~ \tau_{RE} \quad. \eqno(35)$$
We see that time translations simply become rotations in the $\tau_{RE}$ direction which has a period of $\beta=2 \pi$. 
The horizon which was given by $X= \pm T$ is now at $\rho=0$
and world--lines of Rindler observers at constant $\rho$ are circles of radius $\rho$. The proper time of a Rindler observer is given by the arclength of the circular world--line around the origin. The proper energy and temperature measured by these observers are $E_R/\rho$ and $1/2 \pi \rho$ respectively. The entropy is $2 \pi E_R$ and independent of the location of the observer.

Until now, our discussion was completely classical leading to a continuous $E_R$.
However, in quantum mechanics the boundary condition around the periodic $\tau_{RE}$ direction (with period $2 \pi$) requires
$\omega_E 2 \pi=2 \pi n$ where $\omega_E$ is the dimensionless Euclidean angular frequency, i.e. what is usually called the Matsubara frequency in thermal field theory. Thus, the frequency is quantized, $\omega_E=n$. 
Using the definitions $\omega_E=2 \pi f_E$ and $E_R=2 \pi f_E$ (in units with $\hbar=1$) we find that 
$S=2 \pi E_R=2 \pi \omega_E=2 \pi n$. Therefore, the dimensionless Rindler energy is the dimensionless Matsubara frequency which
is an integer and the black hole entropy is quantized in units of $2 \pi$.

We see that (the quantized) black hole entropy arises simply from the quantization of the time evolution along the
$\tau_{RE}$ direction without any relation to the number of Planck area squares on the horizon surface as is commonly assumed in holographic descriptions.
The description is analogous to de Broglie waves in the hydrogen atom with the obvious difference between physical angular motion
around a nucleus and time evolution in Euclidean Rindler space around the black hole. In both cases, the momentum conjugate to
the angular direction, i.e. angular momentum and $2 \pi E_R$ respectively, are quantized due to quantum conditions on 
periodic directions. Quantization arises due to the fact that there has to be an integer number of wavelengths around the periodic (spatial or Euclidean time) direction. Thus we find that the black hole degrees of freedom counted by $2 \pi E_R$ are the number
of ``wavelengths" or more precisely, the number of fundamental frequencies with $n=1$ in the $\tau_{RE}$ direction. 
In the case of de Broglie waves, the underlying physics is described by Quantum Mechanics and the wave is the wave function. Analogously, we hope that the underlying description of
the above picture for the black hole in terms of $E_R$ will be related to the correct theory of quantum gravity. 

It seems from the above discussion that, in some effective way, the black hole entropy can be counted by a periodic quantum mechanical system with a given angular frequency. This effective system somehow has 
entropy $S=2 \pi \omega=2 \pi n$; i.e. it has $e^{2 \pi \omega}$ number of states. In our formalism, thermodynamics arises due to the fact that $d \tau_R/dt=2 \pi T$. As a result of this relation, Eq. (5) which a canonical commutation relation in particle mechanics turns into a thermodynamical statement (i.e. the First Law of Thermodynamics) for the black hole which is an ensemble. 
This can always be done if:

\noindent 
a) $E_R$ is independent of time which is a good approximation for large enough black holes for which Hawking radiation is slow enough to be negligible and 

\noindent
b) time is rescaled linearly with temperature into a dimensionless time i.e. $\tau_R=c T t$ where $c$ is a constant.

An alternative but equivalent way to think about the relation between $E_R$ and entropy is to consider the action
variable $J=\oint p~ dq$ for the above effective, periodic quantum mechanical system[\GOL]. 
Here $q$ and $p$ are a generalized coordinate and its conjugate momentum (such as $\tau_R$ and $E_R$
after the canonical transformation in eq. (5)) and the integral is taken over one period. Before the advent of quantum mechanics, quantization for periodic systems 
was achieved by quantizing the action variable by demanding $J=2 \pi n$ (again in units of $\hbar=1$).
{\footnote5{This is, of course, equivalent to quantization
a la de Broglie by demanding an integer number wavelengths around the periodic direction.}
Since time evolution
in the Euclidean space--time described by the metric in eq. (33) is periodic, using $q=\tau_{RE}$ and $p=E_R$ we get
$$J=\oint E_R d\tau_{RE}= 2 \pi E_R=2 \pi n \quad, \eqno(36)$$
which is exactly the black hole entropy. Thus, the action variable computed in the canonical variables $\tau_{RE},E_R$ is
quantized and equal to the black hole entropy.

One way to motivate this intriguing result is to assume an equality between the free energy $F=E-TS$ of the black hole considered to be a thermal ensemble and the Euclidean action of the effective quantum mechanical system $I_E$, i.e. $I_E=\beta F$ 
where\footnote6{This is different from the Euclidean gravity approach since the $I_E$ is not that of gravity but of the
effective quantum mechanical system.}
$$I_E=-\int dt_E (p_E \dot q_E-H)= -\int p_E dq_E +\beta H \quad. \eqno(37)$$
By comparing $I_E$ to the definition of $F$ we find that if $q_E=\tau_{RE}$ and $p_E=E_R$ we get eq. (36) for the entropy.

Since the effective quantum mechanical system that describes the black hole is periodic, the action variable $J$ is also related 
to the phase space of the system. In the $\tau_{RE},E_R$ plane the phase space is simply a line of height $E_R$ and length $2 \pi$ (due to the periodicity
of $2 \pi$ in the $\tau_{RE}$ direction). If we associate the area under this line with the entropy of the black hole
we get $S=2 \pi E_R$ as required.
After quantization, the phase space is divided into cells of unit area since the minimum area for the canonically conjugate variables $\tau_{RE},E_R$ is determined by the uncertainty relation to be $\Delta \tau_{RE} \Delta E_R \sim 1$. Then, the entropy (or $J$) is simply the number of these cells in phase space.

Until now we refrained from offering a microscopic description of the fundamental degrees of freedom counted by $E_R$. We should note however that in Ref.[\LEN], $E_R$ has been interpreted as the square root of the string oscillator number, 
$E_R=\sqrt{n}$, so that the string entropy (assuming the central charge $c=6$ in all cases)
$S_{string} =2 \pi \sqrt{n}$ gives the black hole entropy. In this picture, the black hole
is described by a very massive, wildly oscillating and very long string at its Hagedorn temperature. The mass of the string is much larger than that of the 
black hole, $m_{string} \sim \sqrt{n} / \ell_s \sim GM^2/\ell_s >> M$ where $M$ is the black hole mass. An asymptotic observer
measures the correct black hole mass due to the gravitational redshift of mass between the near horizon region and asymptotic infinity by the factor $\ell_s/GM$. The Hawking temperature is the redshifted Hagedorn temperature $T_H \sim (\ell_s/GM)(1/\ell_s) \sim 1/GM$.
A string with such a large $n$ is very long, with length $\sim \sqrt{n} \ell_s$ and covers the black hole horizon due to its transverse oscillations. One then gets one string bit per Planck area on the horizon[\UNI].
The string tension is also redshifted to a very small value, $\sim 1/G^2M^2 << 1/\ell_s^2$ so the string does not look like a fundamental one to an asymptotic observer. 

\bigskip
\centerline{\bf 6. Conclusions and Discussion}
\medskip

In this paper, we showed that, in any theory of gravity, the entropy of nonextreme black holes is given by $S=2 \pi E_R$ where $E_R$ is the dimensionless Rindler energy. $E_R$ is an off-shell quantity obtained from the time evolution in the near horizon 
of the black hole. In this respect it is very similar to Hawking temperature. As a result, we get a unified picture of black
hole thermodynamics that is derived solely from the $g_{00}$ component of the metric. 

It is usually assumed that quantum gravity is a holographic theory[\HOL]. The original motivation for this was the Bekenstein--Hawking entropy of black holes which, in General Relativity,  is proportional to horizon area. Even gravitational theories beyond General Relativity seem holographic due to the fact that black hole entropy in these theories is given by an integral over the horizon surface. Unfortunately, the holographic nature of $E_R$ is not manifest since it is not related to the horizon surface, at least not explicitly. {\footnote7{A manifestly holographic formula for $E_R$ as a surface integral (over the horizon) and in terms of the metric can be written. [\SON]}
Moreover. our results indicate that there must be a relation between time evolution in the near horizon region and the number of degrees of freedom on the horizon surface. It would be interesting to make this relation more explicit.
{\footnote8{One can show that $E_R$ is also given by an integral of the local acceleration which only depends on $g_{00}$. [\SON]}

It would be interesting to see if ideas similar to the ones used in this paper can be applied to finite temperature quantum field theories or other thermal systems. 
This may not be so surprising in light of the AdS/CFT correspondence which relates a gravitational theory in the bulk of AdS space with a CFT on its boundary[\ADS]. Consider an AdS Shwarzschild black hole which is dual to a boundary CFT at the Hawking temperature. 
If we compute $E_R$ from the near horizon metric, we find that black hole entropy is correctly given by $S=2 \pi E_R$. On the other hand, the boundary CFT must have the same entropy. Time, which in this case is the dimensionless Rindler time, is common to both the bulk and boundary theories. The bulk and boundary energies are also related.
As a result, there must be a way to formulate entropy in the CFT as the quantity canonically conjugate to a rescaled dimensionless time. Moreover, it seems that this must be true for any thermal QFT and not only those that are holographically dual to gravitational theories. In a generic thermal QFT, after rescaling time by $\tau=c T t$, the dimensionless Euclidean time direction has a periodicity of $c$. Then, in terms of the dimensionless energy $E$ conjugate to $\tau$, the First Law of Thermodynamics becomes $c dE=dS$ leading to $S=cE$. It would be nice to find explicit QFT examples in which entropy is related to a dimensionless energy.

The fundamental, microscopic degrees of freedom that are counted by $E_R$ are obscure. In this paper, we
considered some speculative ideas on these degrees of freedom. The near horizon region of the black hole is a thermal theory with a temperature of $1/2 \pi$. The dimensionless Euclidean (Rindler) time is an angle and
the dimensionless energy is given by $E_R$ (which is the same as the dimensionless angular frequency $\omega$).
Then, as we have shown, the entropy of any black hole is $S=2 \pi E_R$. Quantization leads to an integer $\omega$ due to the periodicity of the Euclidean dimensionless Rindler time. In this description, the fundamental degrees of freedom are units of Euclidean angular frequency. An alternative but related description can be
given in terms of the action variable of an effective one dimensional system which is exactly equal to the entropy. The action variable can be seen as the area in the two dimensional ($\tau_{ER},E_R$) phase space that corresponds to the black hole. Quantization divides the phase space into cells of unit area (in Planck units). In this description, the fundamental degrees of freedom are unit area cells in the phase space and entropy is the number of cells. Unfortunately,  it is not clear how to derive either of these two descriptions of black hole entropy in a theory of quantum gravity such as string theory. Clearly, if the above ideas have any merit, it is important to find out if this is possible.


It seems that a black hole is effectively described by a quantum mechanical periodic system with one degree of freedom whose angular frequency gives the black hole entropy. This is basically a quantum particle in a box where the momentum gives the entropy. The
relation between the quantum particle with one degree of freedom and the black hole which is an ensemble at a finite temperature
is given by the canonical relation in eq. (5) which introduces the temperature into the discussion and leads to thermodynamics.
The canonical transformation described by eq. (5) relates dynamical variables such as $M$ and $t$ to thermodynamical ones
such as $E_R$ and $\tau_R$. It would be helpful to find or build periodic quantum mechanical systems in which the angular frequency
is the entropy. The hope is that examination of such systems will lead to a better understanding of black hole physics and quantum gravity.


\bigskip
\centerline{\bf Acknowledgements}

I would like to thank the Stanford Institute for Theoretical Physics for hospitality.

\vfill

\refout

\end
\bye